\documentclass[pre]{revtex4}

\usepackage{amssymb,amsmath}

\usepackage{color}

\usepackage{graphicx}

\begin{document}
\title{Segregation by thermal diffusion in granular shear flows}
\author{Vicente Garz\'{o}\footnote[1]{Electronic address: vicenteg@unex.es;
URL: http://www.unex.es/eweb/fisteor/vicente/} and Francisco Vega Reyes\footnote[2]
{Electronic address: fvega@unex.es}}
\affiliation{Departamento de F\'{\i}sica, Universidad de Extremadura, E-06071 Badajoz, Spain}
\begin{abstract}

Segregation by thermal diffusion of an intruder immersed in a sheared granular gas is analyzed from the
(inelastic) Boltzmann
equation. Segregation is induced by the presence of a temperature gradient orthogonal to the shear flow plane
and
parallel to gravity.  We show that, like in analogous systems without shear, the segregation
criterion yields a transition between upwards segregation  and
downwards segregation.
The form of the phase diagrams is illustrated in detail showing
that they depend sensitively on the value of gravity relative to the thermal gradient.
Two specific situations are considered: i) absence of gravity, and ii) homogeneous temperature. We
find that
both mechanisms (upwards and downwards segregation) are stronger and more clearly separated when compared with
segregation criteria in systems without shear.

\end{abstract}

\date{\today}
\maketitle

\section{Introduction\label{sec1}}

The understanding of the physical mechanisms involved in segregation phenomena is one of the most important
challenges in the field of granular  matter. Apart from its academic interest, the problem is of central
interest mainly due to its practical relevance in many industrial processes
(powder metallurgy, pharmaceutical pills, glass and paint industries, $\cdots$). In some cases,
it is a desired and useful effect to separate grains of different types (e.g., the separation
of mined ores), while in other situations
the resulting non-uniformity is an undesirable property that can be difficult to control. However, in spite of
its practical importance, the problem is not completely understood yet. This fact has motivated the development
of fundamental theories that provide accurate segregation criteria in the bulk region of the sample \cite{K04}.

One of the most familiar phenomena concerning segregation is the so-called Brazil-nut effect (BNE):
when a binary mixture composed by one large ball and a number of smaller ones is vertically agitated,
usually the intruder (large particle) tends to climb to the top of the sample against gravity
\cite{RSPS87,KJN93,DRC93,CWHB96}. On the other hand, a series of experimental works
\cite{SM98,HQL01} have also observed the reverse buoyancy effect, namely, under certain
conditions the intruder can also sink to the bottom of the granular bed. This effect is
known as the reverse Brazil-nut effect (RBNE). Although several mechanisms have been proposed to
explain the BNE/RBNE transition  \cite{RSPS87,DRC93,KJN93,CWHB96,LCBRD94,SM98,HQL01,MLNJ01,SUKSS06},
the problem is still open. Among the different
competing mechanisms, thermal diffusion becomes the most relevant one when the sample of grains
resembles a granular gas (for example, at large shaking amplitude). In this regime, binary
collisions prevail and kinetic theory can be a quite useful tool to analyze granular
systems. Thermal diffusion (or thermophoresis) in dilute \cite{SGNT06,BRM05,G06bis}
and dense \cite{AW98,JY02,TAH03,G08,GV09,G09} granular mixtures has been a subject
of current interest in the past few years.

A granular gas in rapid flow regime can be achieved by shearing from the boundaries \cite{G03}.
Shearing is in fact as common as shaking in experiments with granular systems. Thus, segregation criteria
for sheared systems are also of interest. Furthermore, experimental works in annular Couette cells \cite{SK08,GD09,MGPSD09,MSD10} have shown that granular materials
segregate by particle size when subjected to shear. Nevertheless, in spite of the relevance of the
problem, much less is known on the theoretical description of segregation
in sheared granular systems.
In effect, to the best of our knowledge, previous theoretical studies \cite{AJ04} on the subject for dense systems have been based on a Chapman-Enskog expansion around Maxwellian distributions at the same temperature for each species \cite{JM89}. But the use of these distributions can only be considered as acceptable for nearly elastic particles where the assumption of the equipartition of energy still holds. Moreover, according to this level of approximation, effects of inelastic collisions appear only through a sink term in the energy balance equation and for this reason the expressions of the Navier-Stokes (NS) transport coefficients for the mixture are the same as those obtained for ordinary gases (elastic collisions). On the other hand, the use of the NS description to analyze segregation in steady granular flows is a serious limitation since the NS theory heavily fails beyond the quasielastic limit \cite{VU09}.
In addition, this is specially true in flows
where viscous heating is exactly balanced by inelastic cooling, for which the granular flow
is inherently non-Newtonian \cite{VSG10}.
For this kind of flows, there is a special case of null temperature gradient, called simple or uniform
shear flow (USF). This flow has received a great deal of attention in the past years and
is the reference case study for granular flows \cite{G03}.

The aim of this paper is to analyze segregation by thermal diffusion in a binary granular mixture under USF in the framework of the inelastic Boltzmann equation. Due to the complexity of the general problem, here we consider the special case in which one of the components is present in tracer concentration. The tracer problem is more amenable to analytical treatment since there are fewer parameters than in a binary system. At a kinetic theory level, in the tracer limit one can assume that the velocity distribution function $f(\textbf{r}, \textbf{v}; t)$ of the granular gas (excess component) obeys the (closed) Boltzmann equation while the velocity distribution function $f_0(\textbf{r}, \textbf{v}; t)$ of the tracer particles satisfies a (linear) Boltzmann-Lorentz equation. The problem is formally equivalent to consider an impurity or intruder immersed in a dilute granular gas, and this will be the terminology used in this paper.

We consider a physical situation where the system (granular gas plus intruder) is in a steady state where {\em weak} spatial gradients of concentration, pressure and temperature coexist with a {\em strong} shear rate, which for the steady USF means strong dissipation \cite{SGD04}. Under these conditions, the resulting diffusion of intruder is anisotropic and, thus, tensorial quantities ($D_{ij}$, $D_{p,ij}$ and $D_{T,ij}$) are required to describe mass transport instead of the conventional scalar transport coefficients \cite{AJ04,G08,GV09}. Explicit expressions for the diffusion tensors $D_{ij}$, $D_{p,ij}$ and $D_{T,ij}$ have been recently obtained \cite{G07bis}
by solving the Boltzmann-Lorentz equation corresponding to the tracer particles by means of a perturbation expansion around a {\em nonequilibrium} sheared state \cite{G02,L06,G06} rather than the (local) equilibrium distribution \cite{CC70}. This is the main new feature of this expansion (in contrast to the usual Chapman-Enskog method) since the reference state retains {\em all} the hydrodynamic orders (NS, Burnet, super-Burnett, $\cdots$) in the shear rate. As a consequence, the different approximations of this expansion are {\em nonlinear} functions of the coefficients of restitution as well as of the parameters of the mixture (masses and sizes).

The knowledge of the diffusion tensors allows us to study segregation by thermal diffusion. On the other hand, due to the anisotropy induced by the shear flow, a thermal diffusion tensor $\boldsymbol{\Lambda}$ is also required to characterize segregation in the different directions. Since in this paper we are interested in steady state conditions, we consider a situation where the temperature gradient is orthogonal to the shear flow plane and parallel to gravity (i.e., $\partial_xT=\partial_yT=0$,
$\partial_z T\neq 0$ and $\partial_x U_y=a\equiv \text{const.}$). In this case, the segregation criterion is obtained from
the thermal diffusion factor $\Lambda_z$, which is given in terms of the generalized
mass transport coefficients $D_{zz}$, $D_{p,zz}$ and $D_{T,zz}$. The use of these generalized non-Newtonian coefficients results in different and more general segregation criteria than those of previous works in sheared systems \cite{AJ04}, which are limited to nearly elastic particles.

Segregation is induced and sustained by both small gravity field and/or temperature gradient.
The signature of $\Lambda_z$ provides a segregation criterion that shows a
transition between upwards and downwards segregation (or BNE and RBNE when the intruder is large) by varying the parameters of the system. In particular, we find
that the form for the upwards/downwards segregation transition depends very sensitively on the value of gravity relative to the thermal gradient in such a way that depending on this one or the other mechanism vastly
predominates in the space parameter. Moreover, our results show differences with those derived \cite{G08,GV09} when the gas is driven by a stochastic thermostat that mimics the effect of a thermal bath \cite{WK96}. These differences lead to interesting results, most notably, we found the upwards/downwards segregation mechanisms are much stronger when shear is input in the system.
This may of course be a signature of more effective segregation process, with a direct impact for applications.

The plan of the paper is as follows. First, the thermal diffusion factor $\Lambda_z$ is defined and
evaluated in
section \ref{sec2} by using a hydrodynamic description.
In section \ref{sec3} we determine the magnitudes needed to calculate $\Lambda_z$ (stress
tensors of gas and intruder and $zz$ elements of the diffusion tensors $D_{ij}$, $D_{p,ij}$ and $D_{T,ij}$).
All these quantities are explicitly obtained after solving the set of (inelastic) Boltzmann equations by
means of the aforementioned Chapman-Enskog-like expansion. The knowledge of the above
quantities yields
$\Lambda_z$ as a function of the parameter space of the problem, namely, the mass and diameter ratios, the two
independent coefficients of restitution for collisions among gas-gas and intruder-gas particles and the reduced gravity (gravity over thermal
gradient). The form of the phase diagrams of segregation
is investigated in section \ref{sec4} by varying the different parameters of the system. In addition,
a comparison with the theoretical results \cite{G06bis} derived when the system is thermalized by a
stochastic thermostat is also carried out. Finally, we briefly discuss the results obtained in this paper in section \ref{sec5}.

\section{Hydrodynamic description for thermal diffusion under shear flow\label{sec2}}

The model system considered is a dilute granular gas of smooth inelastic disks ($d=2$)
or spheres ($d=3$) of mass $m$ and
diameter $\sigma$, plus one intruder or impurity of mass $m_0$ and
diameter $\sigma_0$. The presence of the intruder does not perturb the state of the granular gas and so,
the model system (gas plus intruder) is formally equivalent to a dilute granular binary mixture
in the tracer limit for the impurity component.
Therefore, only gas--gas and intruder--gas particles
collisions need to be taken into account. Collisions are inelastic and characterized
by two independent (constant) coefficients of normal restitution
$\alpha$ and $\alpha_0$, respectively. We assume that the system (gas plus impurity) is in USF.
This flow is characterized by constant
densities $n$, $n_0$ (number densities of gas and intruder  respectively),
uniform granular gas temperature
$T$ and a linear velocity profile $U_{s,x}=ay$ where $a$ is the constant
shear rate. In the USF state the temperature changes in time due to the competition between two (opposite)
mechanisms: on
the one hand, viscous (shear) heating and, on the other hand, energy dissipation in collisions. A steady state
is achieved when both mechanisms cancel each other and the fluid autonomously seeks the temperature at which the
above balance occurs \cite{SGD04}.

The main goal of this paper is to study thermal diffusion of the intruder when the gas is under USF.
We introduce small perturbations to our base state (USF). The perturbations are produced by  weak
gravitational field and small hydrodynamic gradients. These
perturbations give rise to contributions to the mass flux, which can be characterized by generalized
transport coefficients. Therefore, intruder segregation will be determined by the competition between
these two different perturbations and by the relevant parameters values.

The transport properties we need result from a general perturbation of the USF for which
the flow velocity may be expressed as ${\bf U}={\bf U}_s+\delta {\bf U}$ where
$\delta {\bf U}$ is a small perturbation.
Here, ${\bf U}_s={\sf a}\cdot {\bf r}$ with ${\sf a}=a\delta_{ix}\delta_{jy}$. Thus, under these conditions,
the macroscopic balance equations for the system associated with this disturbed USF state are
given by \cite{G07bis}
\begin{equation}
\label{2.0}
\partial_tn+{\bf U}_s\cdot \nabla n+\nabla \cdot (n\delta {\bf U})=0,
\end{equation}
\begin{equation}
\label{2.1}
\partial_tn_0+{\bf U}_s\cdot \nabla n_0=-\nabla \cdot (n_0\delta {\bf U})-
\frac{\nabla \cdot {\bf j}_0}{m_0},
\end{equation}
\begin{equation}
\label{2.2}
\partial_t\delta {\bf U}+{\sf a}\cdot \delta {\bf U}+({\bf U}_s+\delta {\bf U})\cdot \nabla \delta {\bf U}=-
(mn)^{-1}\left(\nabla \cdot {\sf P}-nm{\bf g}\right),
\end{equation}
\begin{equation}
\label{2.3} \frac{d}{2}n\partial_tT+\frac{d}{2}n({\bf U}_s+\delta {\bf U})\cdot \nabla T+aP_{xy}+\nabla \cdot
{\bf q}+{\sf P}:\nabla \delta {\bf U}=-\frac{d}{2}p\zeta,
\end{equation}
where we have assumed that the gravitational field ${\bf g}$ is the only external force. Here,
${\bf g}=-g \hat{{\bf e}}_z$, where $g$ is a positive constant and
$\hat{{\bf e}}_z$ is the unit vector in the positive direction of the $z$ axis.
Moreover,
${\bf j}_0$ is the mass flux of intruder, ${\sf P}$ is
the pressure tensor, ${\bf q}$ is the heat flux and $\zeta$ is the cooling rate associated with
the energy dissipation of collisions between gas particles themselves. It must noted that
the balance equations (\ref{2.0})--(\ref{2.3}) can be exactly obtained from the (inelastic) Boltzmann
equation and they provide the basis for developing a hydrodynamic description of this disturbed USF state.
As usual, to get a hydrodynamic description one has to represent the fluxes ${\bf j}_0$, ${\sf P}$ and
${\bf q}$ as well as the cooling rate $\zeta$ as explicit functionals of the hydrodynamic fields and
their gradients. Once this constitutive equations are determined, the hydrodynamic equations
(\ref{2.0})--(\ref{2.3}) become a closed set of equations for the fields $n$, $n_0$, ${\bf U}$ and $T$.

Thermal diffusion is caused by the relative motion of the components of a mixture due to the presence of
a thermal gradient.
As a consequence of this motion, a steady state is achieved in which the separating
effect arising from the thermal diffusion is balanced by the remixing effect of ordinary diffusion \cite{KCM87}.
We intend to calculate segregation criteria for steady states that weakly deviate from the USF.
Specifically, we are interested in a steady state with $\delta {\bf U}={\bf 0}$ and
with (small) gradients of density, pressure and temperature only along the $z$ axis. These small gradients
are caused by boundary conditions for the temperature and a weak gravitational field.
The hydrodynamic equations (\ref{2.1})--(\ref{2.3})
admit a steady solution for these states.

Let us explain how we will obtain segregation criteria. From an experimental point of view,
the amount of segregation parallel to the thermal gradient
can be characterized by the thermal diffusion factor $\Lambda_z$, which in a steady state is
defined through the relation
\begin{equation}
\label{2.6} \Lambda_z\frac{\partial \ln T}{\partial z} =-\frac{\partial}{\partial z}\ln
\left(\frac{n_0}{n}\right).
\end{equation}
We will consider also that the temperature gradient is directed downwards ($\partial T/\partial z<0$); i.e.
in the same direction as gravity. Thus, when $\Lambda_z >0$ (i.e., $\partial_z\ln (n_0/n)>0$),
the  intruder tends to rise with respect to the gas particles (BNE when the intruder is larger than the particles of the gas)
and when $\Lambda_z <0$ (i.e., $\partial_z\ln
(n_0/n)<0$), the intruder falls with respect to the gas particles (RBNE when the intruder is larger than the particles of the gas).

We obtain now a relation of the type (\ref{2.6}) from the balance equations.
In the steady state with $\delta {\bf U}={\bf 0}$, the mass flux $j_{0,z}$ vanishes according
to the balance equation (\ref{2.1}). To first order in the spatial gradients, the mass flux
$j_{0,z}$ is given by
\begin{equation}
\label{2.9} j_{0,z}=-m_0D_{zz}\frac{\partial x_0}{\partial z}-\frac{m}{T}D_{p,zz}\frac{\partial
p}{\partial z}-\frac{mn}{T}D_{T,zz}\frac{\partial T}{\partial z},
\end{equation}
where $x_0=n_0/n$. Here, $D_{zz}$, $D_{p,zz}$ and $D_{T,zz}$ are the $zz$ elements of the diffusion tensor
$D_{ij}$, the
pressure diffusion tensor $D_{p,ij}$ and the thermal diffusion tensor $D_{T,ij}$, respectively
(As we said, due to the presence of the shear flow, diffusion process is anisotropic and thus
tensorial quantities are required to describe mass transport under USF).
Thus, the condition $j_{0,z}=0$ yields
\begin{equation}
\label{2.10}
\frac{\partial x_0}{\partial z}=-\frac{m}{m_0T}\frac{D_{p,zz}}{D_{zz}}\frac{\partial p}{\partial z}-
\frac{\rho}{m_0T}\frac{D_{T,zz}}{D_{zz}}\frac{\partial T}{\partial z}.
\end{equation}
Momentum balance equation (\ref{2.2}) for our steady state reduces to
\begin{equation}
\label{2.7}
\frac{\partial P_{zz}}{\partial z}=-\rho g,
\end{equation}
where $\rho=mn$ is the mass density of the gas. This equation will allow to express the gradient of
$p$ as a function of the other two gradients and thus, to obtain a relation of the type (\ref{2.6})
from the condition $j_{0,z}=0$.

In the hydrodynamic regime, the pressure tensor $P_{ij}$ has the form \cite{G06,L06}
\begin{equation}
\label{2.8.0}
P_{ij}=pP_{ij}^*(a^*),
\end{equation}
where $p=nT$ is the hydrostatic pressure and $a^*=a/\nu$ is the (reduced) shear rate. Here,
\begin{equation}
\label{2.7.0}
\nu=\frac{\pi^{(d-1)/2}}{\Gamma(d/2)}\frac{8}{d+2}p\sigma^{d-1}\left(mT\right)^{-1/2}
\end{equation}
is an effective collision frequency. According to Eq.\ (\ref{2.8.0}),
the spatial dependence of $P_{zz}$ occurs explicitly
through $p$ and through its dependence on $a^*$. As a consequence,
\begin{eqnarray}
\label{2.7.1}
\frac{\partial P_{zz}}{\partial z}&=&P_{zz}^*\frac{\partial p}{\partial z}+p
\frac{\partial P_{zz}^*}{\partial a^*}\frac{\partial a^*}{\partial z}\nonumber\\
&=&\frac{\partial p}{\partial z}\left(1-a^*\partial_{a^*}\right)P_{zz}^*
+\frac{\partial T}{\partial z}\frac{1}{2}\frac{p}{T}a^*(\partial_{a^*}P_{zz}^*),
\end{eqnarray}
where use has been made of the identity
\begin{equation}
\label{2.7.2}
\frac{\partial a^*}{\partial z}=a^*\left(\frac{1}{2}\partial_z \ln T-\partial_z \ln p\right).
\end{equation}
Using Eq. (\ref{2.7.1}), we can obtain an expression of the gradient of $p$
\begin{equation}
\label{2.8}
\frac{\partial \ln p}{\partial z}=-\frac{\frac{\rho g}{p}+\frac{a^*}{2}(\partial_{a^*}P_{zz}^*)\partial_z
\ln T}
{P_{zz}^*-a^*(\partial_{a^*}P_{zz}^*)}.
\end{equation}
Use of (\ref{2.8}) into (\ref{2.10}) and substitution of Eq.\ (\ref{2.10}) into Eq.\ (\ref{2.6}) finally leads to
\begin{equation}
\label{2.11}
\Lambda_z=\frac{D_{T,zz}^*-\left(P_{zz}^*-a^*(\partial_{a^*}P_{zz}^*)\right)^{-1}
D_{p,zz}^*\left(g^*+\frac{1}{2}a^*(\partial_{a^*}P_{zz}^*)\right)}{D_{zz}^*},
\end{equation}
where we have introduced the reduced coefficients
\begin{equation}
\label{2.12}
D_{zz}^*=\frac{m_0\nu}{p}D_{zz}, \quad D_{p,zz}^*=\frac{m\nu}{Tx_0}D_{p,zz}, \quad
D_{T,zz}^*=\frac{m\nu}{Tx_0}D_{T,zz}
\end{equation}
and
\begin{equation}
\label{2.13} g^*=\frac{\rho g}{n\left(\frac{\partial T}{\partial z}\right)}<0
\end{equation}
is a dimensionless parameter measuring the gravity relative to the thermal gradient.
This quantity measures the competition between these two mechanisms ($g$ and $\partial_z T$) on segregation.

The condition $\Lambda_z=0$ provides the criterion for the upwards/downwards segregation transition.
Since the diffusion coefficient $D_{zz}^*$ is positive (as will be shown later), according to Eq.\
(\ref{2.11}), the condition $\Lambda_z=0$ implies
\begin{equation}
\label{2.14}
\left(P_{zz}^*-a^*(\partial_{a^*}P_{zz}^*)\right)D_{T,zz}^*=
D_{p,zz}^*\left(g^*+\frac{1}{2}a^*(\partial_{a^*}P_{zz}^*)\right).
\end{equation}
This equation delimits the upwards and downwards segregation regimes in a granular gas driven by
shear flow. The explicit form of the pressure tensor and the diffusion coefficients relevant for
this problem were calculated in previous works \cite{L06,G06,SGD04,G07bis}.
We shortly explain in the next section the procedure followed to obtain them.

\section{Boltzmann kinetic theory}
\label{sec3}

We adopt now a kinetic theory point of view and start from the set of Boltzmann kinetic equations for the
system (gas plus intruder). In this description, all the macroscopic properties of interest of the system are
determined from the one-particle distribution function of the gas $f(\textbf{r}, \textbf{v}; t)$ and
the one-particle distribution function of the impurity $f_0(\textbf{r}, \textbf{v}; t)$. While the
time evolution of the distribution function $f(\textbf{r}, \textbf{v}; t)$  is given by the (closed) inelastic Boltzmann
equation \cite{GS95,BDS97}, the distribution function $f_0(\textbf{r}, \textbf{v}; t)$  obeys the (linear)
Boltzmann-Lorentz equation.

In this paper we are interested in a flow that differs {\em slightly} from the steady USF \cite{VSG10,SGD04}. Under these conditions, the kinetic equations for $f$ and $f_0$ have been recently solved \cite{G07bis,G06,G02,L06} by means of a Chapman-Enskog-like expansion around local shear flow distributions. Therefore, we look for solutions of the form
\begin{equation}
\label{n1}
f=f^{(0)}+f^{(1)}+\cdots, \quad f=f_0^{(0)}+f_0^{(1)}+\cdots,
\end{equation}
where the reference zeroth-order distribution functions $f^{(0)}$ and $f_0^{(0)}$ are the corresponding local versions of the USF distributions of the gas and the intruder, respectively. According to this perturbation scheme, the successive approximations $f^{(k)}$ and $f_0^{(k)}$ are of order $k$ in the spatial gradients of concentration $x_0$, pressure $p$ and temperature $T$ but \emph{retain} all the hydrodynamic orders in the shear rate $a$. As said in the Introduction, this is the main new ingredient of this expansion with respect to the conventional Chapman-Enskog method \cite{CC70}. This feature allows us to deploy for the first time a segregation theory based on non-Newtonian hydrodynamics for granular sheared systems.

\begin{figure}
\includegraphics[width=0.5 \columnwidth,angle=0]{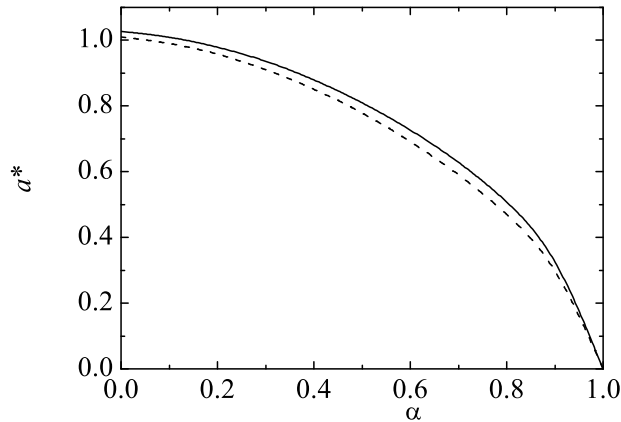}
\caption{Plot of the reduced shear rate $a^*$ as a function of the coefficient of restitution $\alpha$ for $d=2$ (dashed line) and $d=3$ (solid line).  \label{fign0}}
\end{figure}

The rheological properties of the reference states $f^{(0)}$ and $f_0^{(0)}$ are related to the pressure tensors ${\sf P}^{(0)}$ and ${\sf P}_0^{(0)}$, respectively. They are defined as
\begin{equation}
\label{n2}
P_{ij}^{(0)}=\int \; d{\bf v} m V_iV_j  f^{(0)}({\bf V}), \quad P_{0,ij}^{(0)}=\int \; d{\bf v} m V_iV_j  f_0^{(0)}({\bf V}),
\end{equation}
where ${\bf V}={\bf v}-{\bf U}_s$ is the peculiar velocity. In addition, the steady state condition in the USF problem requires that the viscous heating term $a^*P_{xy}^*$ is exactly compensated for by the collisional cooling term $\zeta^*$ \cite{VSG10,SGD04}. In that case, the (reduced) shear rate $a^*$ and the coefficient of restitution $\alpha$ are not independent parameters but they are coupled by the relation
\begin{equation}
\label{n3}
a^*P_{xy}^*=-\frac{d}{2}\zeta^*,
\end{equation}
where $P_{ij}^*=P_{ij}^{(0)}/p$ and $\zeta^*=\zeta/\nu$. Equation (\ref{n3}) shows the intrinsic connection between the shear field and dissipation in the system. In fact, in the elastic limit ($\alpha=1$), $\zeta^*=0$ and the (reduced) shear rate vanishes.
The explicit expressions for the (reduced) pressure tensor $P_{ij}^*$, $P_{0,ij}^*=P_{0,ij}^{(0)}/x_0p$ and their derivatives with respect to
$a^*$ in the steady USF were obtained in Refs.\ \cite{L06,G06,SGD04}. They are displayed in the Appendix \ref{appA} for the sake of completeness. Since $\alpha\leq 1$ the range of (reduced) shear rates is defined, according to Eq.\ (\ref{a4}), in the interval $0\leq a^*\leq (3+2d)\sqrt{(d+2)/(32d(d+1))}$. Thus, for hard disks ($d=2$), $0\leq a^*\lesssim 1.01$ while $0\leq a^*\lesssim 1.03$ for hard spheres ($d=3$). This clearly shows that non-Newtonian effects become important as the dissipation increases. To illustrate this non-Newtonian behavior, Fig.\ \ref{fign0} shows $a^*$ versus $\alpha$ for hard disks and spheres. It is apparent that the (reduced) shear rate (which gives the steady granular temperature) is a \emph{nonlinear} function of $\alpha$ in contrast to previous analysis \cite{AJ04} carried out for nearly elastic particles where $a^{*2}\propto 1-\alpha$.

\begin{figure}
\includegraphics[width=0.5 \columnwidth,angle=0]{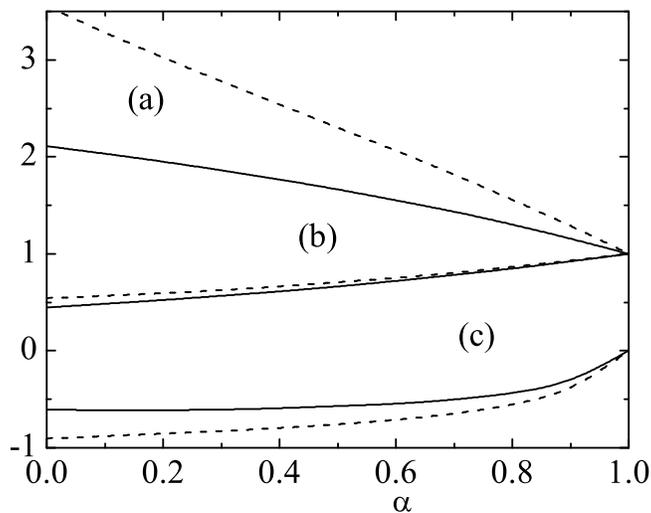}
\caption{Dependence of the diagonal elements (a) $P_{xx}^*$ and $P_{0,xx}^*$, (b) $P_{yy}^*$ and $P_{0,yy}^*$, and the off-diagonal elements (c) $P_{xy}^*$ and $P_{0,xy}^*$ on the (common) coefficient of restitution $\alpha=\alpha_0$ in the three-dimensional case for $\sigma_0/\sigma=2$ and $m_0/m=4$. The solid lines correspond to the elements $P_{ij}^*$ of the granular gas  while the dashed lines are the elements $P_{0,ij}^*$ of the intruder particles.  \label{fign1}}
\end{figure}

To complement the results shown in Fig.\ \ref{fign0}, the dependence of the relevant elements of $P_{ij}^*$ and $P_{0,ij}^*$ on $\alpha$ is plotted in Fig.\ \ref{fign1} for the system $d=3$, $\sigma_0/\sigma=2$ and $m_0/m=4$. As expected, the transport properties in the steady USF state are inherently different from those of the NS description [which means here nearly elastic particles due to Eq.\ (\ref{n3})]. A non-Newtonian signal of this behavior is the existence of the normal stress differences in the shear flow plane. It must remarked that, although the expressions of $P_{ij}^*$ and $P_{0,ij}^*$ have been obtained by an approximate solution based on Grad's method, their $\alpha$-dependence compare quite well with Monte Carlo simulations even for strong dissipation \cite{BRM97,MG02,MG02bis,GM03}.

Once the rheological properties of the gas and the intruder are well characterized, the diffusion coefficients $D_{zz}$, $D_{p,zz}$ and $D_{T,zz}$ can be obtained to first order in the expansion.
The expressions of these generalized coefficients were derived in Ref.\ \cite{G07bis}. While the coefficient $D_{zz}$
decouples from the other two, the coefficients $D_{p,zz}$ and $D_{T,zz}$ obey a set of coupled equations.
These equations are solved in the Appendix \ref{appB}. The results show that $D_{zz}>0$ while
$D_{p,zz}$ and $D_{T,zz}$ have not a definite signature.


Before exploring the dependence of the parameter space on the form of the phase diagrams, it is instructive to consider certain limit situations. For example,
there is no segregation when the intruder and the particles of gas are
mechanically equivalent
($m_0=m$, $\sigma_0=\sigma$, and $\alpha_0=\alpha$). This is consistent with the results derived
in the Appendix \ref{appB} since in this limit
$D_{p,zz}=D_{T,zz}=0$ [the
right hand side of eqs.\ (\ref{b3}) and (\ref{b4}) vanish since $P_{zz}^*=P_{0,zz}^*$] and so, $\Lambda_z=0$
for all values of $\alpha$.
Another reference situation is the elastic limit [$\alpha=
\alpha_0=1$, which implies $a^*=0$ in the steady state condition (\ref{n3})].
In this limit case, $P_{zz}^*=1$ and the diffusion coefficients
behave as
\begin{equation}
\label{3.1}
D_{zz}\to\frac{p}{m_0\omega_0}, \quad D_{p,zz}\to \frac{Tx_0}{m\omega_0}\left(1-\frac{m_0}{m}\right),
\quad D_{T,zz}\to 0,
\end{equation}
where
\begin{equation}
\label{3.1.0}
\omega_0=\frac{4\pi^{(d-1)/2}}{d\Gamma(d/2)}n\left(\frac{\sigma+\sigma_0}{2}\right)^{d-1}\sqrt{\frac{2mT}{m_0(m+m_0)}}
\end{equation}
is a (positive) collision frequency.
Consequently, for elastic collisions the segregation criterion
(\ref{2.14}) becomes
\begin{equation}
\label{3.2}
g^*\left(1-\frac{m_0}{m}\right)=0.
\end{equation}
In the absence of gravity, Eq.\ (\ref{3.2}) holds trivially, and so the intruder does not segregate. On the
other hand, when $|g^*|\neq 0$ the solution to (\ref{3.2}) is simply $m_0=m$, namely,
segregation is predicted for particles that differ in mass, no matter what their diameters may be \cite{G06bis}.
For inelastic systems, in general the criterion
(\ref{2.14}) is rather complicated since it involves all the parameter space of the problem.
\begin{figure}
\includegraphics[width=0.5 \columnwidth,angle=0]{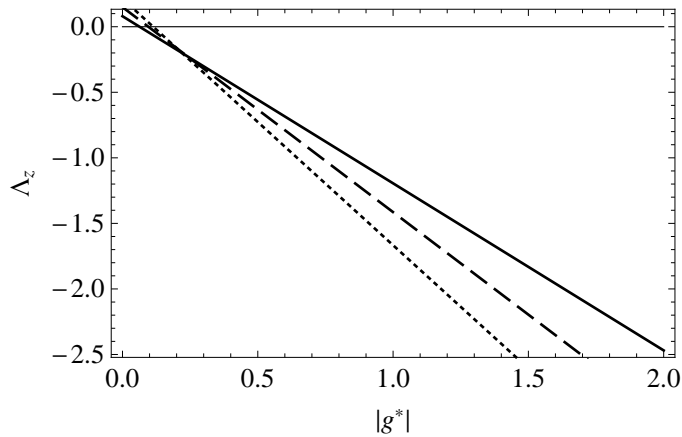}
\caption{Plot of thermal diffusion factor $\Lambda_z$ \emph{versus} the (reduced) gravity $|g^*|$ for
for inelastic hard spheres ($d=3$) in the case $m_0/m=\sigma_0/\sigma=2$.
Three different values of the common ($\alpha=\alpha_0$) coefficient of restitution
$\alpha$ have been considered: $\alpha=0.9$ (solid line), $\alpha=0.8$ (dashed line), and
$\alpha=0.7$ (dotted line). \label{fig1}}
\end{figure}
\begin{figure}
\includegraphics[width=0.5 \columnwidth,angle=0]{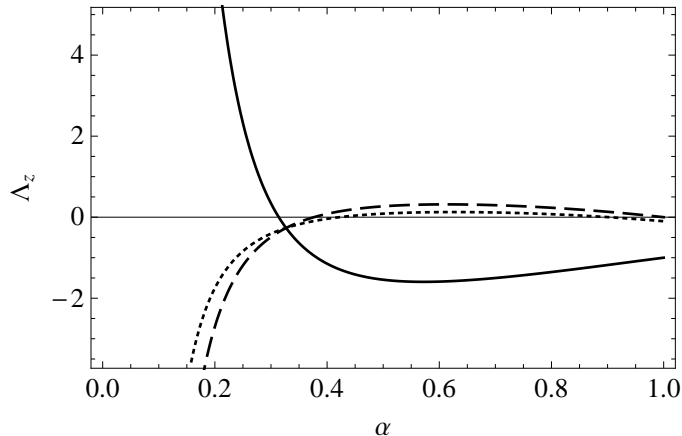}
\caption{Plot of thermal diffusion factor $\Lambda_z$ \emph{versus}  common ($\alpha=\alpha_0$) coefficient of restitution
$\alpha$ for inelastic hard spheres ($d=3$) in the case $\sigma_0/\sigma=1$ and $m_0/m=2$.
Three different values of the reduced gravity $|g^*|$
have been considered: $|g^*|=1$ (solid line), $|g^*|=0.1$ (dotted line), and
$|g^*|=0$ (dashed line). \label{fig2}}
\end{figure}

Equation (\ref{2.11}) clearly shows that $\Lambda_z$
is a linear function of the reduced gravity $|g^*|$. This is illustrated in figure \ref{fig1} where thermal
diffusion is plotted as a function of $|g^*|$ for $m_0/m=\sigma_0/\sigma=2$ and several values of a common coefficient of restitution ($\alpha=\alpha_0$). For the case represented here, downwards segregation is
dominant, except for quite small values of $|g^*|$
(i.e., there is a critical value $|g^*_c|$ such that a transition upwards segregation $\Rightarrow$ downwards segregation
occurs for $|g^*|>|g^*_c|$). It is to be noticed that the opposite behavior is present in the
segregation results for dense systems driven by a stochastic thermostat \cite{G08,G09}.
Figure \ref{fig2} shows the $\alpha$-dependence of the thermal diffusion factor for
different values of the reduced gravity $|g^*|$. We observe that the presence of gravity
changes dramatically the regions of positive and negative $\Lambda_z$. In particular,
for the case considered here and for not quite small values of $\alpha$, we observe that the
downwards segregation region increases with the reduced gravity.

\section{Phase diagrams for segregation}
\label{sec4}

As it can be noticed from Eq.\ (\ref{2.11}) and from the Appendix A, the thermal diffusion
factor $\Lambda_z$ depends on many parameters: the dimensionless gravity $g^*$, the mass
ratio $m_0/m$, the size ratio $\sigma_0/\sigma$, and the coefficients of restitution $\alpha$ and $\alpha_0$.
Given that the parameter space is fivefold, for simplicity, we take the simplest case of common coefficient
of restitution $\alpha=\alpha_0$. This reduces the parameter space to four quantities.

The zero contour of $\Lambda_z$ separates regions of positive (upwards segregation) and negative (downwards segregation)
$\Lambda_z$. Points lying
on the zero contour correspond to values of the parameters of the system for which the intruder does not
segregate.
According to Eq.\ (\ref{2.14}), segregation is sustained by both gravity and the thermal gradient. The combined
effect of both $g$ and $\partial_zT$ on thermal diffusion is through the dimensionless gravity $g^*<0$ defined
by Eq.\ (\ref{2.13}). Although our previous perturbation analysis \cite{G07bis} assumes that the external
field is of the same order of magnitude as $\partial_zT$, it is instructive to separate the influence of
each one of the terms appearing in Eq.º (\ref{2.14}) on segregation.
Thus, two specific limit situations will be
considered first in the next subsections: (a) absence of gravity ($g=0$), and
(b) homogeneous temperature
($\partial_zT=0$).

\begin{figure}
\includegraphics[width=0.5 \columnwidth,angle=0]{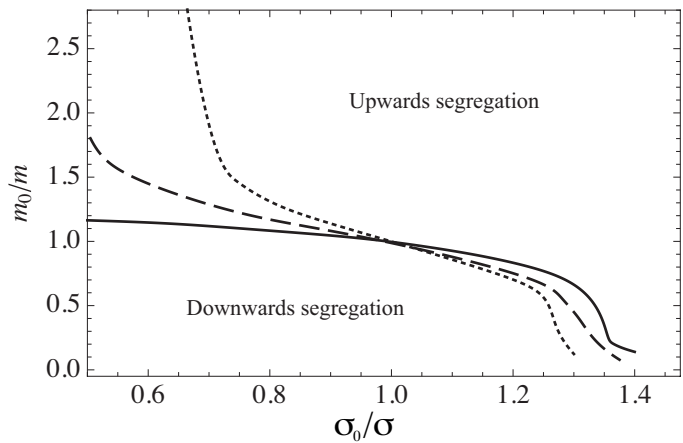}
\caption{Phase diagram for segregation for inelastic hard spheres ($d=3$) in the absence of
gravity ($|g^*|=0$). Three different values of the (common) coefficient of restitution
$\alpha$ have been considered: $\alpha=0.9$ (solid line), $\alpha=0.8$ (dashed line), and
$\alpha=0.7$ (dotted line). Points above the curve correspond to $\Lambda_z>0$ (upwards segregation) while points
below the curve correspond to $\Lambda_z<0$ (downwards segregation).  \label{fig3}}
\end{figure}

\subsection{Absence of gravity ($|g^*|=0$)}

In this case, segregation of the intruder is due only to the presence of the thermal gradient.
Under these conditions, $|g^*|\to 0$ and Eq.\ (\ref{2.14})
reduces to
\begin{equation}
\label{4.1}
P_{zz}^*D_{T,zz}^*=a^*(\partial_{a^*}P_{zz}^*)\left(
D_{T,zz}^*+\frac{1}{2}D_{p,zz}^*\right).
\end{equation}
Of course, this condition is trivially satisfied in the elastic case (for which
$\alpha=1$, $a^*=0$, and $D_{T,zz}^*=0$).
For inelastic systems ($\alpha \neq 1$), the influence of each term in Eq.\ (\ref{4.1}) is still intricate due
to the presence of shear flow. As an illustration, figure \ref{fig3} shows the phase diagram in the
$\left\{m_0/m, \sigma_0/\sigma\right\}$-plane for three different values of the coefficient of restitution
$\alpha$. As expected, when $m_0/m=\sigma_0/\sigma=1$,
the species are indistinguishable, and so $\Lambda_z=0$.
Consequently, all zero contours of $\Lambda_z$ pass through the
point $(1,1)$. It is apparent that, for $\sigma_0<\sigma$,
the main effect of dissipation (or equivalently, the reduced shear rate $a^*$) is
to reduce the size of the upwards segregation region while
the opposite happens when the intruder is larger than the particles of the gas. In the latter case, the
influence of dissipation on the phase diagram is much smaller than in the former case
(when $\sigma_0<\sigma$) so that all the curves tend to collapse in a common one
for sufficiently large values of the
size ratio $\sigma_0/\sigma$.  We also observe that in the zero gravity limit large intruders
will in general tend to move towards hotter regions, since the upwards segregation
is dominant and occupies most of the parameter space: in fact, for large intruders ($\sigma_0/\sigma>1$),
upwards segregation is the only active mechanism for $\sigma_0/\sigma\gtrsim1.4$.
This is to be put in contrast with the case of no shear (vibrated systems in dense systems),
where segregation tends to be predominantly of the downwards segregation type as the
intruders get larger \cite{G09}, just the opposite behavior seen here for sheared granular gases.

\begin{figure}
\includegraphics[width=0.5 \columnwidth,angle=0]{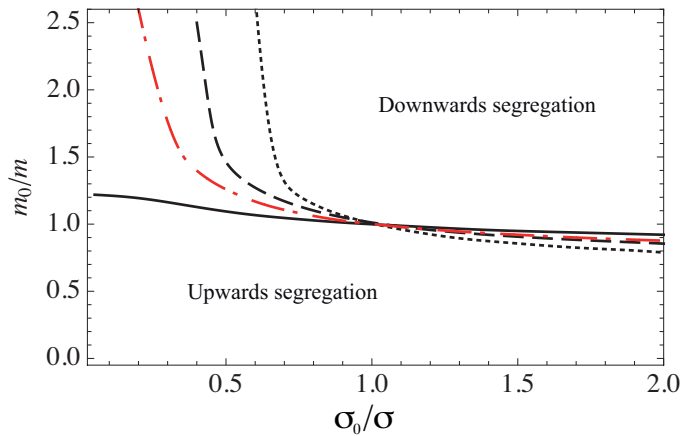}
\caption{(color online) Segregation phase diagram for inelastic hard spheres ($d=3$) in the absence of
thermal gradient ($|g^*|\to \infty$). Three different values of the (common) coefficient of restitution
$\alpha$ have been considered: $\alpha=0.9$ (solid line), $\alpha=0.8$ (dashed line), and
$\alpha=0.7$ (dotted line). Points above the curve correspond to $\Lambda_z<0$ (downwards segregation) while points
below the curve correspond to $\Lambda_z>0$ (upwards segregation). The dashed-dotted line refers to the results
obtained when the gas (for $\alpha=0.8$) is driven by a stochastic external force. \label{fig4}}
\end{figure}

\subsection{Thermalized systems ($\partial_zT=0$)}

Let us consider now a system with both negligible temperature and mole fraction gradients
($\partial_zT\to 0, \partial_zx_0\to 0$).
Thus, the segregation of the intruders
is only driven by gravity. This situation (gravity dominates over thermal gradient) can be achieved in
vibrated or sheared systems in computer simulations and real experiments \cite{HQL01,BEKR03,SBKR05,WHP01}.
In this case ($|g^*|\to \infty$), the sign of $\Lambda_z$ is the same as that of the pressure
diffusion coefficient $D_{p,zz}^*$ and so, the criterion (\ref{2.14}) becomes simply
\begin{equation}
\label{4.2}
D_{p,zz}^*=0.
\end{equation}
In the elastic case, $D_{p,zz}^*\propto m-m_0$ and so, the segregation criterion is $m_0=m$. For inelastic
gases, the dependence of $D_{p,zz}^*$ on the parameter space is quite complex.

As we  said in the Introduction, previous results for thermal diffusion have been obtained when the gas
is driven by means of a stochastic external force (thermostat) that mimics the effect of a thermal bath.
This external driving method is usually employed in computer simulations \cite{thermostat} to compensate for
cooling effects associated with the inelasticity of collisions. Under these conditions,
an explicit expression
for the thermal diffusion factor based on the NS transport coefficients has been recently obtained
\cite{G06bis}. This expression for $\Lambda_z$ leads to the segregation criterion \cite{G06bis}
\begin{equation}
\label{4.3}
\frac{T_0}{T}=\frac{m_0}{m},
\end{equation}
where the temperature ratio is determined from the condition $\zeta_0T_0/m_0=\zeta T/m$.
The segregation criterion
(\ref{4.3}) compares well with molecular dynamics simulation results for the case of the steady state of an
open vibrated granular system in the absence of macroscopic fluxes \cite{BRM05}.
As expected, it is clear that the segregation criterion (\ref{4.3})
(obtained from the NS description) differs from the one derived here for sheared gases (see Appendix
\ref{appB}). The discrepancies between (\ref{4.2}) and (\ref{4.3}) are a direct consequence of the inherent non-Newtonian features of the USF state not present at the NS level.

A typical phase diagram for thermalized sheared systems delineating the regimes between upwards and downwards segregation is plotted in figure \ref{fig4}. The results obtained from the relation (\ref{4.3}) (stochastic driving thermostat) for
$\alpha=0.8$ are also shown for comparison.
In spite of the differences between the criteria (\ref{4.2}) and (\ref{4.3}),
the shape of the phase diagram obtained from both relations agrees very well
for large size ratios. In particular, when $\sigma_0/\sigma<1$, the main effect
of inelasticity is to reduce the size of the downwards segregation region, while the opposite happens when $\sigma_0/\sigma>1$.
On the other hand, the effect of dissipation on the form of the phase diagrams is much more important when
$\sigma_0<\sigma$ than when $\sigma_0>\sigma$. Moreover,
comparison between figures \ref{fig3} and \ref{fig4} clearly shows that the presence of gravity changes
dramatically the form of the phase diagram since the regions of positive and negative $\Lambda_z$ are
interchanged.

In the same way as it happens for $|g^*|=0$, there is one clearly predominant segregation mechanism,
the downwards segregation this time, being the only active mechanism for $m_0/m>1$ if $\sigma_0/\sigma>1$. Interestingly, the
downwards segregation  is not present for small but heavy intruders if the granular gas is sufficiently inelastic.
This last situation is in contrast with the vibrated systems since there the downwards segregation still occurs in this region \cite{G09}.

\begin{figure}
\includegraphics[width=0.5 \columnwidth,angle=0]{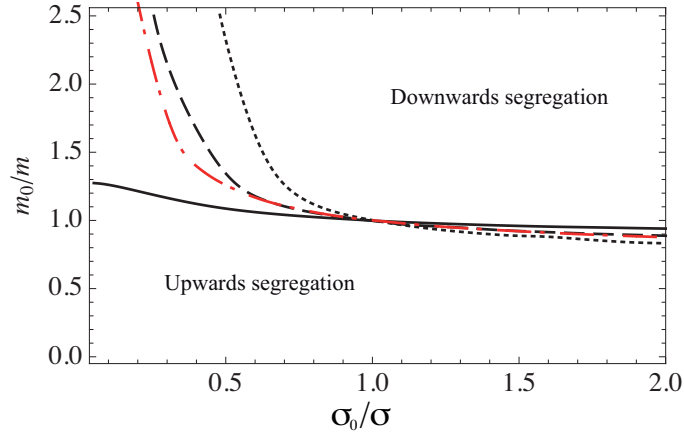}
\caption{(color online) Segregation phase diagram for inelastic hard spheres ($d=3$) in the case
$|g^*|=1$. Three different values of the (common) coefficient of restitution
$\alpha$ have been considered: $\alpha=0.9$ (solid line), $\alpha=0.8$ (dashed line), and
$\alpha=0.7$ (dotted line). Points above the curve correspond to $\Lambda_z<0$ (downwards segregation) while points
below the curve correspond to $\Lambda_z>0$ (upwards segregation). The dashed-dotted line refers to the results
obtained when the gas (for $\alpha=0.8$) is driven by a stochastic external force. \label{fig5}}
\end{figure}
\begin{figure}
\includegraphics[width=0.5 \columnwidth,angle=0]{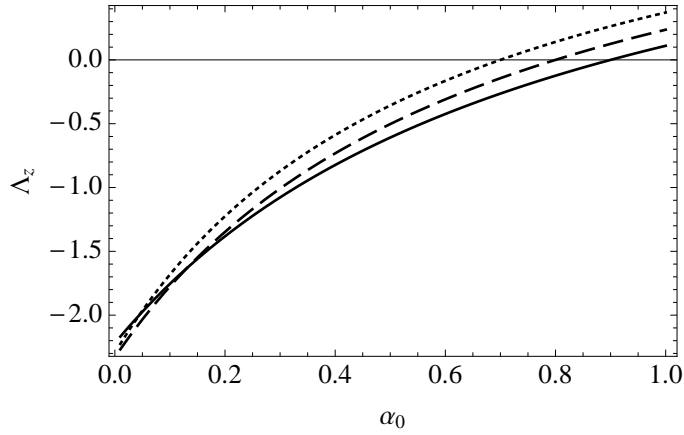}
\caption{Plot of thermal diffusion factor $\Lambda_z$ \emph{versus} the coefficient of restitution
$\alpha_0$ for inelastic hard spheres ($d=3$) for the system $\sigma_0/\sigma=m_0/m=1$ when $|g^*|=1$.
Three different values of the coefficient of restitution $\alpha$
have been considered: $\alpha=0.9$ (solid line), $\alpha=0.8$ (dashed line), and
$\alpha=0.7$ (dotted line).  \label{fig6}}
\end{figure}

\subsection{General case}

We analyze the form of the phase diagrams for finite values of the reduced gravity $|g^*|$. Figure
\ref{fig5} shows the phase diagram when $|g^*|=1$ (gravity comparable to the thermal gradient) for the same
cases as considered in the previous figures. In general, the form of the phase diagram for $|g^*|=1$ is quite
similar to the one obtained when $|g^*|=\infty$; i.e., $|g^*|=1$ is already sufficiently large to show
the high gravity segregation behaviour (similar to that of the thermalized systems ). Comparison with the stochastic driving results shows now better agreement with the sheared results than those observed for thermalized systems (see figure \ref{fig4}).

All previous graphs have been obtained by assuming a common coefficient of restitution. Therefore, it is
worth studying the (pure) effect on inelasticity on segregation in shear flows. In order to
illustrate this case, we consider the case $m_0=m$, and $\sigma_0=\sigma$ with $|g^*|=1$, corresponding to a
system of particles that differ by their coefficients of restitution only. Clearly, when all the coefficients
of restitution are equal ($\alpha=\alpha_0$), the thermal diffusion factor vanishes ($\Lambda_z=0$)
and so, as expected
the intruder does not segregate. Figure \ref{fig6} presents plots of $\Lambda_z$ versus $\alpha_0$ for
different values of $\alpha$. When the coefficients of restitution are different from each other, thermal
diffusion is different from zero so that segregation in the presence of a temperature gradient can therefore
occur as a consequence of inelasticity only. This effect is also present when the mixtures are vertically
vibrated \cite{SGNT06,B08}. But, \emph{again} contrary to what happens in the vibrated systems,
in the sheared system the upwards segregation region increases for increasing inelasticity.

Therefore, we have seen that in most cases the qualitative behavior of the segregation criteria for
the sheared system is rather different from the criteria for vibrated systems. Moreover the shearing
has the effect of clearly favoring just one segregation mechanism: upwards segregation for $|g|<|g_c|$ and downwards segregation for $|g|>|g_c|$.

\section{Discussion}
\label{sec5}

The problem of segregation of an intruder in a sheared granular gas has been addressed in this paper.
The relative motion of the intruder with respect to the particles of the gas is caused by the combined effect
of gravity and a temperature gradient. In this case, thermal diffusion (or thermophoresis in its single-particle
manifestation) forces large or massive particles to move down temperature gradients \cite{GR83}. Under these conditions, the amount of segregation parallel to the thermal gradient can be measured
by the thermal diffusion factor.
However, the analysis of thermal diffusion  in a \emph{strongly} shearing gas is an intricate problem basically due to the anisotropy induced in the system by the shear flow. For this reason, in general a thermal diffusion tensor is needed to describe the segregation process in the different directions. In this paper, for the sake of simplicity we have assumed that the temperature gradient is orthogonal to the shear flow plane.

Under the above conditions, the thermal diffusion factor $\Lambda_z$ [defined by Eq.\ (\ref{2.6})] has been
obtained in a steady state where \emph{weak} spatial gradients of concentration, pressure and temperature
directed along the vertical direction (parallel to gravity) coexist with a \emph{strong} (constant)
shear rate $a$. Our approach to determine the thermal diffusion factor follows two
complementary routes. First,
by using a hydrodynamic description, $\Lambda_z$ has been expressed in terms of the pressure tensor
$P_{ij}$ of the gas and the transport coefficients $D_{zz}$, $D_{p,zz}$, and $D_{T,zz}$ associated with the mass
flux of the intruder. Second, we have adopted a kinetic theory point of view and have considered some previous
results \cite{G02,G07bis} obtained by solving the Boltzmann kinetic equation for the gas and the Boltzmann-Lorentz equation for the intruder by means of a Chapman-
Enskog-like type of expansion. This allows us to compute the thermal diffusion factor as a function of the mass
and size ratios, the coefficients of restitution for gas-gas and intruder-gas collisions and the reduced gravity
$g^*=\rho g/n\partial_z T<0$. Once the explicit form of $\Lambda_z$ is known, the condition $\Lambda_z=0$
provides the segregation criterion for the transition upwards segregation (regions of positive $\Lambda_z$)$\Leftrightarrow$
downwards segregation (regions of negative $\Lambda_z$). This criterion is given by equation (\ref{2.14}) in terms of the
(reduced) $zz$-element of the pressure tensor $P_{zz}^*$ and the diffusion transport coefficients
$D_{zz}^*$, $D_{p,zz}^*$, and $D_{T,zz}^*$. In order to make the paper self-contained, the complete expressions
for all the above quantities have been displayed in the Appendices \ref{appA} and \ref{appB}.

Some previous works \cite{AW98,AJ04} on thermal diffusion segregation in sheared systems
have been based on the elastic NS transport coefficients. These coefficients are obtained
by the Chapman-Enskog expansion around the local equilibrium distribution. Thus, their expressions
are limited to \emph{nearly} elastic systems in the USF problem. The present study takes the (local)
shear flow distributions for the gas particles and the intruder as reference states in the Chapman-Enskog
method so that, the corresponding transport coefficients derived from this expansion retain \emph{all} the
hydrodynamic orders (NS, Burnett, super-Burnett, $\cdots$) in the shear rate. As a consequence, since the coefficients of restitution and the shear rate
are not independent parameters in the steady USF [see Eqs. (\ref{n3}) and/or (\ref{a4})],
our segregation criterion goes beyond the weak-dissipation limit. This is perhaps the main new
added value with respect to previous theoretical results \cite{AW98,AJ04}.
Moreover, our theory also takes into account the influence of both thermal gradients and gravity through the reduced gravity $g^*$.

In order to illustrate the form of the phase diagrams in the mass and size ratio plane, a (common)
coefficient of restitution $\alpha=\alpha_0$ has been
assumed. Two specific situations have been mainly
studied: $|g^*|=0$ (absence of gravity) and $|g^*|\to \infty$ (homogeneous temperature).
The shear field has the neat effect of selecting more strongly the upwards segregation mechanism: we have found that upwards segregation predominates very strongly for small gravity $|g^*|$ (is in fact the only active mechanism for large and
heavy intruders if gravity is small). These results are in contrast with segregation in the
absence of shear, where there is no such a strong predominance of the upwards segregation mechanism
for small gravities \cite{G09}. Regarding the impact of the inelasticity of collisions on segregation, the results show that
the influence of dissipation on thermal diffusion is more important when the thermal gradient dominates over
gravity ($g^*=0$) than in the opposite limit ($|g^*|\to \infty$), especially when the intruder is smaller than the particles of the gas. In addition, in general the effect of inelasticity on segregation mechanisms is less sensitive when the diameter of the intruder is greater than the particles of the gas. We also observe that the influence of gravity on the form of phase diagrams is quite important: while upwards segregation is dominant for small $g^*$, downwards segregation dominates if $g^*$ is large.
To complement the previous study, we have also considered the
case of identical mass and size for the intruder and gas particles but different coefficients of restitution ($\alpha\neq \alpha_0$). As it happens for vibrated granular mixtures \cite{SGNT06,B08},
the intruder can segregate
in a sheared granular gas on the basis of differences in their \emph{inelastic} properties.

The results derived here for thermal diffusion have been obtained in the tracer limit $x_0\to 0$. This limit
precludes the possibility of
analyzing the influence of composition on the thermal diffusion factor $\Lambda_z$,
where previous results for dilute gases \cite{G06bis} have shown that the effect of $x_0$
on $\Lambda_z$ can be significant in many situations. The study on the dependence of thermal
diffusion on composition is an interesting open problem.
It is also apparent that the results presented here
are relevant to make a comparison with numerical simulations and/or experiments. In this context, we hope that this paper
stimulates the performance of such simulations/experiments to check the relevance of kinetic theory to describe
thermal diffusion segregation under shear flow. We plan to work along the above lines in the near future.

\acknowledgments

This work has been supported by the Ministerio de Educaci\'on y Ciencia (Spain) through grant No.
FIS2007-60977, partially financed by
FEDER funds and by the Junta de Extremadura (Spain) through Grant No. GRU10158.

\appendix
\section{Pressure tensor of the gas and of the impurity in the steady USF}
\label{appA}

The non-zero elements of the (reduced) pressure tensor of the gas $P_{ij}^*=P_{ij}/p$ are given by \cite{SGD04}
\begin{equation}
\label{a1} P_{yy}^*=P_{zz}^*=\cdots=P_{dd}^*=\frac{d+1+(d-1)\alpha} {2d+3-3\alpha}, \quad
P_{xx}^*=d-(d-1)P_{yy}^*,
\end{equation}
\begin{equation}
\label{a2} P_{xy}^*=-4d\frac{d+1+(d-1)\alpha} {(1+\alpha)(2d+3-3\alpha)^2}a^*,
\end{equation}
where the relationship between the reduced shear rate $a^*=a/\nu$ [where $\nu$ is defined by
Eq.\ (\ref{2.7.0})] and the coefficient of restitution $\alpha$ is
\begin{equation}
\label{a4}
a^{*}(\alpha)=\sqrt{\frac{d+2}{32d} \frac{(1+\alpha)(2d+3-3\alpha)^2(1-\alpha^2)}{ d+1+(d-1)\alpha}}.
\end{equation}

Moreover, the (reduced) cooling rate $\zeta^*=\zeta/\nu$ is
\begin{equation}
\label{a5} \zeta^*=\frac{d+2}{4d}(1-\alpha^2).
\end{equation}
The derivative of $P_{zz}^*$ with respect to $a^*$ is  \cite{G07bis,L06,G06}
\begin{equation}
\label{a6} \frac{\partial P_{zz}^*}{\partial a^*}=\frac{\partial P_{yy}^*}{\partial a^*}=4 P_{zz}^*\frac{a^*
\Delta+P_{xy}^*}{2a^{*2} \Delta +d( 2\beta+\zeta^*)},
\end{equation}
where
\begin{equation}
\label{a7} \beta=\frac{1+\alpha}{2}\left[1-\frac{d-1}{2d}(1-\alpha)\right],
\end{equation}
and $\Delta\equiv \left(\partial P_{xy}^*/\partial a^*\right)$ is the real root of the cubic equation
\begin{equation}
\label{a8} 2 a^{*4} \Delta^3+4da^{*2}(\zeta^*+\beta)\Delta^2+\frac{d^2}{2}(7\zeta^{*2}
+14\zeta^*\beta+4\beta^2)\Delta+d^2\beta(\zeta^*+\beta)^{-2}( 2\beta^2-2\zeta^{*2}-\beta\zeta^*)=0.
\end{equation}

Apart from the pressure tensor of the gas, another relevant transport property is the partial pressure tensor
${\sf P}_0$ of the impurity. This quantity along with its derivative with respect to $a^*$ is also needed
to compute the diffusion coefficients. The elements of the reduced pressure tensor $P_{0,ij}^*=P_{0,ij}/n_0T$
are \cite{G02}
\begin{equation}
\label{a9} P_{0,yy}^*=P_{0,zz}^*=\cdots=P_{0,dd}^*=-\frac{F+HP_{yy}^*}{G},
\end{equation}
\begin{equation}
\label{a10} P_{0,xy}^*=\frac{a^*P_{0,yy}^*-HP_{xy}^*}{G},
\end{equation}
\begin{equation}
\label{a11} P_{0,xx}^*=d\gamma-(d-1)P_{0,yy}^*,
\end{equation}
where $\gamma=T_0/T$ is the temperature ratio and
\begin{equation}
\label{a12} F=\frac{\sqrt{2}}{2d}\left(\frac{\overline{\sigma}}{\sigma}
\right)^{d-1}M_{0}\left(\frac{1+\theta}{\theta^3}\right)^{1/2}(1+\alpha_{0})
\left[1+\frac{M}{2}(d-1)(1+\theta)(1+\alpha_{0})\right],
\end{equation}
\begin{eqnarray}
\label{a13} G&=&-\frac{\sqrt{2}}{4d}\left(\frac{\overline{\sigma}}{\sigma}
\right)^{d-1}M\left(\frac{1}{\theta(1+\theta)}\right)^{1/2}
(1+\alpha_{0})\nonumber\\
& & \times \left\{2[(d+2)\theta+d+3]-3M (1+\theta)(1+\alpha_{0})\right\},
\end{eqnarray}
\begin{equation}
\label{a14} H=\frac{\sqrt{2}}{4d}\left(\frac{\overline{\sigma}}{\sigma}
\right)^{d-1}M_{0}\left(\frac{1}{\theta(1+\theta)}\right)^{1/2}
(1+\alpha_{0})\left[3M(1+\theta)(1+\alpha_{0})-2\right].
\end{equation}
Here, $\overline{\sigma}=(\sigma+\sigma_0)/2$, $\theta=m_0T/mT_0$ is the mean-square velocity of
the gas particles relative to
that of the impurity, $M=m/(m+m_0)$ and $M_0=1-M=m_0/(m+m_0)$.
The temperature ratio $\gamma$ is determined from the condition
\begin{equation}
\label{a15} \gamma=\frac{\zeta^*P_{0,xy}^*}{\zeta_0^*P_{xy}^*},
\end{equation}
where the ``cooling rate'' $\zeta_0^*=\zeta_0/\nu$ for the impurity is given by
\begin{equation}
\label{a16} \zeta_0^*=\frac{(d+2)\sqrt{2}}{2d}\left(\frac{\overline{\sigma}}{\sigma}
\right)^{d-1}M\left(\frac{1+\theta}{\theta}\right)^{1/2}(1+\alpha_{0})
\left[1-\frac{M}{2}(1+\theta)(1+\alpha_{0})\right].
\end{equation}

The derivative of the elements of $P_{0,ij}^*$ with respect to $a^*$ is more involved. They can be written as
\cite{G07bis}
\begin{equation}
\label{a17} \frac{\partial P_{0,yy}^*}{\partial a^*}=\frac{\partial P_{0,zz}^*}{\partial a^*}=\frac{\chi
P_{0,yy}^*+\left(Y'+X_0'P_{0,yy}^*+X'P_{yy}^*\right)(\partial\gamma/\partial a^*)+X (\partial
P_{yy}^*/\partial a^*)} {\frac{1}{2}a^*\chi-X_0},
\end{equation}
\begin{eqnarray}
\label{a18} \frac{\partial P_{0,xy}^*}{\partial
a^*}&=&\left(\frac{1}{2}a^*\chi-X_0\right)^{-1}\left\{ \chi
P_{0,xy}^*+\left(X_0'P_{0,xy}^*+X'P_{xy}^*\right)(\partial\gamma/\partial a^*)\right.\nonumber\\
& & \left.- \left[P_{0,yy}^*+a^*(\partial P_{0,yy}^*/\partial a^*)\right]+X\Delta\right\} ,
\end{eqnarray}
where $\chi=(2/d)(P_{xy}^*+a^*\Delta)$ and
\begin{equation}
\label{a18b} Y=\frac{d+2}{2\sqrt{2}d}\left(\frac{\overline{\sigma}}{\sigma}\right)^{d-1}M_0
(1+\alpha_0)\left(\frac{1+\theta}{\theta}\right)^{3/2} \left[\frac{\lambda_{0}}{d+2}+\frac{d}{d+3}M
(1+\alpha_{0})\right],
\end{equation}
\begin{equation}
\label{a19} X_0=-\frac{d+2}{\sqrt{2}d}\left(\frac{\overline{\sigma}}{\sigma}\right)^{d-1}
M_0(1+\alpha_0)\left[\theta(1+\theta)\right]^{-1/2} \left[
1+\frac{(d+3)}{2(d+2)}\frac{1+\theta}{\theta}\lambda_{0}\right] \gamma^{-1},
\end{equation}
\begin{equation}
\label{a20} X=\frac{d+2}{\sqrt{2}d}\left(\frac{\overline{\sigma}}{\sigma}\right)^{d-1}
M_0(1+\alpha_0)\left[\theta(1+\theta)\right]^{-1/2} \left[1-\frac{(d+3)}{2(d+2)}(1+\theta)\lambda_{0}\right],
\end{equation}
with
\begin{equation}
\label{a21} \lambda_{0}=\frac{2}{1+\theta}-\frac{3}{d+3}M_0(1+\alpha_{0}).
\end{equation}
Moreover, $Y'\equiv (\partial Y/\partial \gamma)$, $X'\equiv (\partial X/\partial \gamma)$,
$X_0'\equiv (\partial
Y/\partial \gamma)$, $\zeta_0'\equiv (\partial \zeta_0^*/\partial \gamma)$ and $\partial \gamma/\partial a^*=
\Lambda_1/\Lambda_2$ where
\begin{eqnarray}
\label{a22} \Lambda_1&=&d\left(\frac{1}{2}a^*\chi-X_0\right)\left\{
\left(\frac{1}{2}a^*\chi-X_0\right)\left(\chi\gamma-\frac{2}{d} P_{0,xy}^*\right)-\frac{2}{d}a^*\left(\chi P_{0,xy}^*-P_{0,yy}^*+X \Delta\right)\right\}\nonumber\\
& &+2a^{*2}\left[\chi P_{0,yy}^*+X(\partial P_{yy}^*/\partial a^*)\right],
\end{eqnarray}
\begin{eqnarray}
\label{a23} \Lambda_2&=&d\left(\frac{1}{2}a^*\chi-X_0\right)\left[
\left(\frac{1}{2}a^*\chi-X_0\right)\left(\zeta_0^*+\frac{1}{2}\chi+\gamma\zeta_0'\right)\right. \nonumber\\
& & \left.+\frac{2}{d}a^*\left(X_0'P_{0,xy}^*+X'P_{yy}^*\right)\right]-
2a^{*2}\left(Y'+X_0'P_{0,yy}^*+X'P_{yy}^*\right).
\end{eqnarray}

\section{Diffusion coefficients $D_{zz}^*$, $D_{p,zz}^*$ and $D_{T,zz}^*$
\label{appB}}

In this Appendix we give the explicit expressions of the tensors $D_{zz}^*$, $D_{p,zz}^*$ and $D_{T,zz}^*$.
The diffusion coefficient $D_{zz}^*$ is given by \cite{G07bis,G02}
\begin{equation}
\label{b1}
D_{zz}^*=\frac{P_{0,zz}^*}{\Omega_{zz}^*},
\end{equation}
where
\begin{eqnarray}
\label{b2} \Omega_{zz}^*&=&\frac{\sqrt{2}}{4d}\left( \frac{\overline{\sigma}}{\sigma}\right)^{d-1}M
(1+\alpha_{0})\left[(1+\theta)\theta\right]^{-1/2} \gamma P_{0,zz}^{*-1} \left\{(d+2)(1+\theta)\right.
\nonumber\\
& & \left. +\theta\left(P_{zz}^*-1\right)+\left[(d+3)+(d+2)\theta\right]
\left(\gamma^{-1}P_{0,zz}^*-1\right)\right\}.
\end{eqnarray}
In the region of parameter space explored, $\Omega_{zz}^*>0$ so that the diffusion coefficient $D_{zz}^*$ is
always positive.

The pressure diffusion $D_{p,zz}^*$ and the thermal diffusion $D_{T,zz}^*$ coefficients
are coupled. They obey the set of algebraic equations \cite{G07bis}
\begin{eqnarray}
\label{b3}  \left[\frac{2}{d}a^*(P_{xy}^{*}-a^*\Delta)+2\zeta^{*}-
\Omega_{zz}^*\right]D_{p,zz}^*&-&
\left(\frac{2}{d}a^{*2}\Delta-\zeta^{*} \right)D_{T,zz}^*
\nonumber\\
&=&(1-a^*\partial_{a^*})
\left(\frac{m_0}{m}P_{zz}^*-P_{0,zz}^*\right),
\end{eqnarray}
\begin{eqnarray}
\label{b4}  \left[\frac{2}{d
}a^*(P_{xy}^{*}+\frac{1}{2}a^*\Delta)+\frac{1}{2}\zeta^{*}-\Omega_{zz}^*\right]D_{T,zz}^*
&+&\left(\frac{a^{*2}}{d}\Delta-\frac{1}{2}\zeta^{*} \right)D_{p,zz}^*\nonumber\\
&=&\frac{1}{2}a^*\partial_{a^*}
\left(\frac{m_0}{m}P_{zz}^*-P_{0,zz}^*\right),
\end{eqnarray}
where $\Delta$ is the real root of (\ref{a8}) and the expressions of the nonzero elements of
the pressure tensors $P_{ij}^*$ and $P_{0,ij}^*$ and their
derivatives with respect to $a^*$ are given in the Appendix \ref{appA}.
The solution to the set of equations (\ref{b3}) and (\ref{b4}) is elementary and provides
the explicit expressions of $D_{T,zz}^*$ and $D_{p,zz}^*$ in terms of $m_0/m$, $\sigma_0/\sigma$,
$\alpha$ and $\alpha_0$. The results show that while $D_{zz}^*>0$, the coefficients $D_{p,zz}^*$ and
$D_{T,zz}^*$ have not a definite sign.

\end{document}